\documentclass[12pt,english]{iopart} 

\usepackage[usenames,dvipsnames]{color}
\usepackage[normalem]{ulem}
\usepackage{graphicx}

\expandafter\let\csname equation*\endcsname\relax
\expandafter\let\csname endequation*\endcsname\relax

\usepackage{amsmath}
\usepackage{amssymb,amsthm}
\usepackage{amsfonts}
\usepackage{color}
\usepackage{graphics}
\usepackage{epsfig}
\usepackage{bbm}
\usepackage{pstricks}
\usepackage{subfigure}
\usepackage{bm}
\usepackage{bbold}
\usepackage{cite}

\newcommand{\eins}{\mathbbm{1}}

\newcommand{\exv}[1]{{\langle{#1}\rangle}}

\newcommand{\bra}[1]{\langle #1|}
\newcommand{\ket}[1]{|#1\rangle}
\newcommand{\braket}[2]{\langle #1|#2\rangle}
\newcommand{\ketbra}[2]{|#1\rangle\!\langle#2|}
\newcommand{\trace}{\operatorname{Tr}}

\newcommand{\vr}{\varrho}

\begin{document}

\title{The structure of ultrafine entanglement witnesses}

\author{Mariami Gachechiladze, Nikolai Wyderka, and Otfried G\"uhne}

\address{Naturwissenschaftlich-Technische Fakult\"at,
Universit\"at Siegen,\\
Walter-Flex-Stra{\ss}e~3,
57068 Siegen, Germany}


\date{\today}

\begin{abstract} 
An entanglement witness is an observable with the property that a negative expectation
value signals the presence of entanglement. The question arises how a witness can be 
improved if the expectation value of a second observable is known, and methods for 
doing this have recently been discussed as so-called ultrafine entanglement witnesses. 
We present several results on the characterization of entanglement given the expectation
values of two observables. First, we explain that this problem can naturally be tackled
with the method of the Legendre transformation, leading even to a quantification of entanglement. 
Second, we present necessary and sufficient conditions that two product observables 
are able to detect entanglement. Finally, we explain some fallacies in the original 
construction of ultrafine entanglement witnesses [F. Shahandeh {\it et al.}, 
Phys. Rev. Lett. {\bf 118}, 110502 (2017)].
\end{abstract}

\section{Introduction}
Entanglement is a central phenomenon in quantum information processing 
and entanglement witnesses provide some of the most 
effective tools for detecting it \cite{hororeview, Guehne2009}. Mathematically, 
an entanglement
witness is a Hermitean operator $W$, such that $\trace(\sigma W) \geq 0$ holds
for every separable state $\sigma \in \mathcal{S}_{\rm sep}$. Consequently, 
measuring a negative expectation value $\trace(\vr W)<0$ implies that the 
state $\vr$ is entangled. 

Any entanglement witness can be written as  
\begin{equation}
\label{eq:basicwitness}
W = g_s \mathbbm{1} - L,
\end{equation}
where $L$ is the observable actually to be measured in an 
experiment and $g_s$ characterizes the relation of $L$ to
separability. It is defined as the maximal value that $L$
can attain on separable states, $g_s = \sup_\sigma 
\{ \trace(\sigma L)|\sigma\in \mathcal{S}_{\rm sep} \}$.
The physical interpretation of the witness is then clear: If one measures
an expectation value $l=\exv{L}$ larger than $g_s$, the state
must be entangled and $\exv{W} <0$. 

Some remarks are in order.
First, since the set of separable states is defined as the convex hull 
of all pure product states, the state $\sigma_{\rm opt}$ where 
the expression $\{ \trace(\sigma L)|\sigma\in \mathcal{S}_{\rm sep} \}$
is maximal is a pure product state, which simplifies the computation 
of $g_s.$ Second, for many possible $L$, the computation of $g_s$ can be carried out 
analytically, for instance if $L=\ketbra{\psi}{\psi}$
is a projector onto a pure state \cite{Guehne2009}. Finally, it can happen, of course, that the 
observable $L$ is not useful for entanglement detection at all. This is the 
case, if the maximal eigenvalue of $L$ coincides with $g_s$ and consequently
no state can have an expectation value $\exv{L}$ that exceeds the maximal value for
separable states.

What changes, if in addition to $l=\exv{L}$ the expectation value $c=\exv{C}$ of a
second observable $C$ is known? From the knowledge of the two observables, one can
determine the expectation value of $X= \alpha C + \beta L$ for any values of 
$\alpha, \beta.$ Consequently, any witness of the type
\begin{equation}
\label{eq:parameterwitness}
W(\alpha, \beta) = g_s(\alpha, \beta) \mathbbm{1} - X
\end{equation}
can be evaluated. From the convexity of the set of separable states it also follows
that any state whose entanglement can be proved from the knowledge of $l=\exv{L}$
and $c=\exv{C}$ must be detected by the witness from Eq.~(\ref{eq:parameterwitness})
for some $\alpha, \beta$.

A different approach, called ultrafine entanglement witnessing (UEW) was recently 
introduced \cite{Shahandeh2017, Erratum2017} and further developed \cite{newuew}. 
Here, one starts from Eq.~(\ref{eq:basicwitness})
and asks how $g_s$ can be changed due to the knowledge of $c=\exv{C}$. In fact, one
can then define
\begin{equation}
\label{eq:constrainedg}
g_s = \sup_\sigma
\{ \trace(\sigma L)|\sigma\in \mathcal{S}_{\rm sep} \mbox{ and } \trace(\sigma C)= c\}.
\end{equation}
Of course, evaluating this is more complicated, and the question arises how one
can characterize the optimal $\sigma_{\rm opt; c}$ which is the separable state 
obeying ${\rm Tr}( C \sigma_{\rm opt; c}) = c$ and maximizing the expectation 
value of $L$. 

A second question is, whether one can derive conditions on $L$ and $C$ which 
guarantee that the knowledge of ${\rm Tr}(\varrho {C})$ improves the capability 
of $L$ to detect entanglement. In an experimental setting, one typically considers
observables which are easy to implement, so one may take $L=L_A \otimes L_B $
and $C=C_A \otimes C_B $ to be product observables. In this case, $l=\exv{L}$  (or $c=\exv{C}$)
alone is clearly not sufficient to detect entanglement, so the question arises what 
conditions $C$ and $L$ have to fulfill in order to detect entanglement together. So 
far, a conjecture has been presented \cite{Shahandeh2017}, but its rigorous derivation 
remained elusive.

In this paper, we present several results on the characterization of entanglement
from the expectation values of two observables. In Section~II we start with the 
problem in Eq.~(\ref{eq:parameterwitness}). We explain how the problem can be solved
using the method of the Legendre transformation. We stress that this approach 
is not new \cite{Guehne2007, eisert2007}, but we present an analytical example
useful for our later discussion. In Section~III we study the question, which 
conditions $L=L_A \otimes L_B$ and $C=C_A \otimes C_B $ have to fulfill in order 
to be able to detect entanglement. We solve the problem for two qubits and present 
extensions, as well as interesting counterexamples, for higher dimensions. In Section~IV 
we critically discuss the results of Ref.~\cite{Shahandeh2017} in some detail. Some 
errors in this reference have been already corrected in an erratum \cite{Erratum2017},  
but it is instructive to study precisely the fallacies.   
Finally, we conclude and discuss some open problems  for further research.

\section{Legendre transformation}

In this section, we consider the general task of characterizing entanglement 
from two known expectation values. More precisely, given the two expectation
values $c = \trace{(\vr C)}$ and $l = \trace{(\vr L)}$ of the observables $C$ 
and $L$, we want to find a lower bound on the value of some entanglement measure 
$E(\vr)$ of the state $\vr$. The method we explain uses the Legendre transformation.
It is not new and has been introduced in Refs.~\cite{Guehne2007, eisert2007}.

The connection to UEW introduced in Ref.~\cite{Shahandeh2017} is simple: 
The method of Legendre transformations gives a non-trivial bound on some
faithful entanglement measure for the states compatible with $c$ and $l$, 
if and only if these observables are useful for UEW. UEWs, however, can 
only certify entanglement, while the method of Legendre transformation provides 
a systematic way to estimate entanglement quantitatively. In
addition, the Legendre transformation treats the observables $C$ and $L$ in a
symmetric manner.

\subsection{The general method}
We need to compute the minimal value of $E(\vr)$ over all states compatible with 
the observed data,
\begin{equation}
\varepsilon(c, l) = 
\inf_\vr \{E(\vr) \,\vert\, \trace{(\vr C)} = c \mbox{ and } \trace{(\vr L)} = l\}.
\end{equation}
If the entanglement measure $E(\vr)$ is convex, then this is a convex function in $c$ 
and $l$. As such, it can be characterized as the supremum over all affine functions below 
it. Therefore, given $c$ and $l$, we would like to find the smallest constant 
$k\in \mathbb{R}$, such that
\begin{equation}
\label{eq:inequality1}
\varepsilon(c,l) \geq \alpha c+ \beta l - k
\end{equation}
for arbitrary $\alpha$ and $\beta \in \mathbb{R}$. Rewriting 
Eq.~(\ref{eq:inequality1}), we obtain
\begin{align}
k := \hat{E}(X)
= & \sup_{c,l}\{\alpha c+ \beta l - \varepsilon(c,l)\} \nonumber \\
= & \sup_{\varrho}\{\alpha \trace{(\varrho C)}+ \beta \trace{(\varrho L)} - E(\varrho)\},
\label{eq:kdef}
\end{align}
which is the definition of $\hat{E}$ as the Legendre transform of the entanglement 
measure $E$, evaluated for the operator
\begin{equation}
X = \alpha C + \beta L.
\end{equation}
The value of $k$ can in turn be used to obtain the supremum over all slopes 
$\alpha$ and $\beta$:
\begin{equation}
\label{eq:lowerbound}
\varepsilon(c,l)=\sup_{\alpha,\beta}\{ \alpha c+ \beta l -\hat{E}(X) \}.
\end{equation}
This is itself a Legendre transform of $\hat{E}(X)$ and gives a lower bound 
on the entanglement measure $E(\vr)$ from the values $c$ and $l$. It follows from 
convex geometry that this lower bound is optimal, in the sense that there is one
state with the values $c$ and $l$ having the entanglement $E(\vr)=\varepsilon(c,l)$.
Also, it should be noted that there is a practical difference between Eq.~(\ref{eq:kdef}) 
and Eq.~(\ref{eq:lowerbound}): For obtaining a valid lower bound on $E(\vr)$ one needs 
the global optimum in the maximization in Eq.~(\ref{eq:kdef}). In Eq.~(\ref{eq:lowerbound}), 
however, any pair of values $\alpha, \beta$ gives already a valid lower bound. 

\subsection{A concrete example}

Whether one can analytically evaluate Eqs.~(\ref{eq:kdef}, \ref{eq:lowerbound}) 
depends on the entanglement measure $E$ and the specific form of $C$ and $L$. For measures 
that are defined via the convex roof construction,
\begin{equation}
E(\vr) = \inf_{p_i, \ket{\psi_i}} \sum_i p_i E(\ket{\psi_i}),
\end{equation}
with $\vr=\sum_i p_i \ketbra{\psi_i}{\psi_i}$, the Legendre transform can be 
evaluated by optimizing over pure states only \cite{Guehne2007}:
\begin{equation}
E(X)=\sup_{\ket{\psi}} \{ \bra{\psi}X\ket{\psi} - E(\ket{\psi})\}.
\end{equation}
Here, we concentrate on the geometric measure of entanglement $E_G$ \cite{weigeometric} 
defined for pure states as one minus the maximal overlap with product states, 
\begin{equation}
E_G(\ket{\psi}) = 1-\sup_{\ket{\phi}=\ket{a}\ket{b}\ket{c}\dots}
\left|\left\langle \phi|\psi\right\rangle \right|^{2},
\end{equation}
and for mixed states via a convex roof construction. Thus, we need to evaluate
\begin{equation}
\label{eq:optimizeme}
    E_G(X) = \sup_{\ket{\psi}} \sup_{\ket{\phi}=\ket{a}\ket{b}\ket{c}\dots}
    \{\bra{\psi}(X+\ketbra{\phi}{\phi})\ket{\psi} - 1\}.
\end{equation}
This optimization can be done in practice numerically in an efficient manner 
\cite{Guehne2007}. Here, however, we consider a specific case of $X$ where 
analytical derivations can be made. We choose
\begin{equation}\label{eq:lowerboundXXZZ}
C = \sigma_z \otimes \sigma_z, \quad L = \sigma_x \otimes \sigma_x.
\end{equation}
The operator $X$ accordingly is then diagonal in the Bell basis with eigenvalues 
$(\alpha+\beta,\alpha-\beta,-\alpha+\beta, -\alpha-\beta)$. For such 
operators, the Legendre transformation of the geometric measure is given by 
\cite{Guehne2008}
\footnote{Note that the formula in Ref.~\cite{Guehne2007} is formulated 
as an upper bound on the Legendre transform, but for the special case of two-qubit 
Bell states equality holds. This is due to the fact that for any pair of Bell 
states we can find a product vector having an overlap of 1/2 with both. Finally, 
it should be added that Eq.~(28) in Ref.~\cite{Guehne2008} contains a typo.}
\begin{equation}
E_G(X) = \frac{\lambda_1 + \lambda_2-1}{2} + \frac12\sqrt{(\lambda_1-\lambda_2)^2 +1},
\end{equation}
where $\lambda_1$ denotes the largest, and $\lambda_2$ denotes the second-largest 
eigenvalue of X.

For $c = \trace{(\vr C)}$ and $l = \trace{(\vr L)}$ we can assume without loss of 
generality that  $c\geq l\geq 0$, as this can be achieved for the given observables 
by local unitary transformations that do not alter the entanglement. In addition, we
have $c\leq 1$. It is easy to 
see that the higher the values of $c$ and $l$, the more entangled the state is. From this 
it follows that in Eq.~(\ref{eq:lowerbound}) the interesting case is if $\alpha$ and 
$\beta$ both have positive values. 

\begin{figure}[t]
    \centering
    \includegraphics[width=0.5\columnwidth]{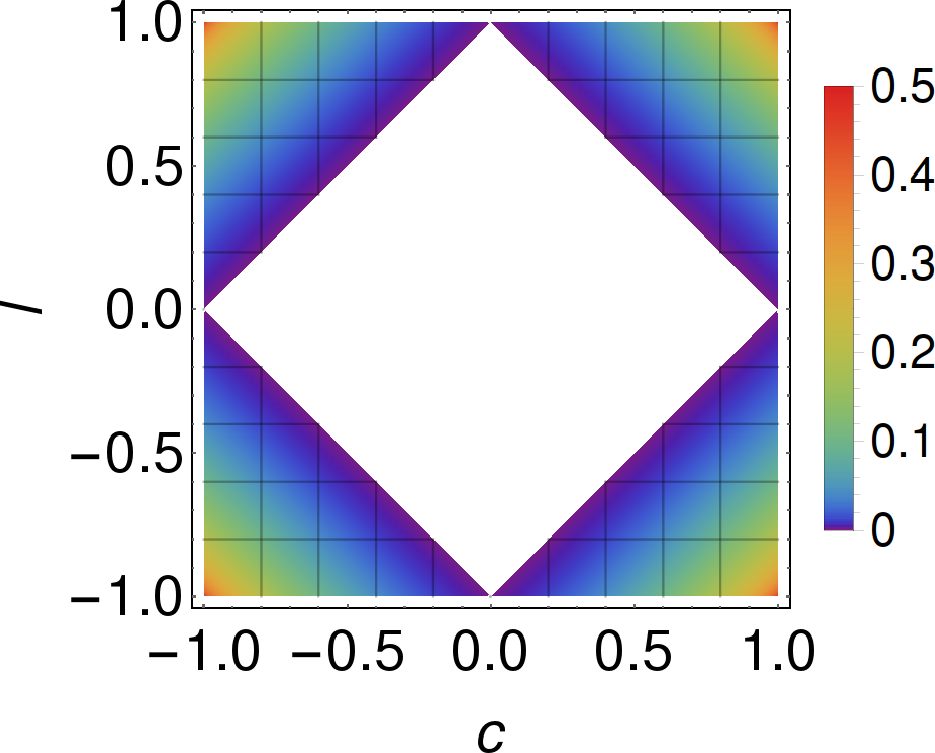}
    \caption{Lower bounds on the geometric measure of entanglement based on 
    the expectation values $c=\trace(\varrho \sigma_z \otimes \sigma_z)$ and 
    $l=\trace(\varrho \sigma_x \otimes \sigma_x)$. The white inner region 
    corresponds to separable states.} 
    \label{fig:geombounds}
\end{figure}

We have to distinguish between two cases, $\alpha \geq \beta \geq 0$ 
and $\beta \geq \alpha \geq 0$. First, assume that $\alpha\geq\beta\geq 0$, 
then 
\begin{align}
\label{eq:derivation1}
    \varepsilon(c,l) 
    = & \sup_{\alpha,\beta}\bigg[\alpha(l-1)+\frac{1}{2}+
    \beta c-\sqrt{\beta^2+\frac{1}{4}} \bigg].
\end{align}
Taking the partial derivative with respect to $\alpha$ we obtain 
$\partial_{\alpha}\varepsilon(c,l)=(l-1)\leq 0$, therefore $\alpha$ 
must be chosen as small as possible, i.e.,~$\alpha=\beta$. Inserting 
this and taking the partial derivative with respect to $\beta$ we 
find
\begin{equation}
\partial_{\beta}\varepsilon(c,l) 
= 
(c + l- 1)-\frac{\beta}{\sqrt{\beta^2+\frac{1}{4}}} \overset{!}{=} 0,
\end{equation}
and consequently $\beta={(c+l-1)}/{[2\sqrt{1-(c+l-1)^2}]}$. This yields the final result:
\begin{equation}
\label{eq:resultonXXZ}
     E_G  \geq \varepsilon(c,l) = \frac12\left( 1-\sqrt{1-(c+l-1)^2} \right).
\end{equation}
Considering the second case, $\beta\geq\alpha\geq 0$, the second largest 
eigenvalue changes from $\alpha-\beta$ to $\beta-\alpha$. The function 
to maximize is essentially the same as before, but with $\alpha$ and $\beta$
swapped. Therefore, the solution is the same and Eq.~(\ref{eq:resultonXXZ})
also holds. The corresponding bounds on the geometric measure are depicted 
in Fig.~\ref{fig:geombounds}.

\subsection{Necessary and sufficient criterion for two qubits}
In this case, we can directly formulate the main result:

\noindent
{\bf Proposition 1.}
{\it Consider a two-qubit system and product operators $C=C_A \otimes C_B$ and 
$L=L_A \otimes L_B$. $C$ and $L$ can be used for entanglement detection, if and 
only if $[C_A,L_A]\neq 0$ and $[C_B,L_B]\neq 0$.}  

{\it Proof.}
One direction is trivial and valid for any dimension: If $[C_A,L_A]=0$, 
then $C$ and $L$ cannot be used for entanglement detection. The reason 
is that in this case Alice is effectively performing only a single 
measurement $M_A$. So, any possible linear combination $X = \alpha C +\beta L$ 
can be evaluated from the statistics of a product measurement 
$M_A \otimes M_B$, (where $M_B$ may depend on $\alpha,\beta$) and for such 
measurements the probabilities can always be mimicked by a separable state. 
A similar reasoning holds if $[C_B,L_B]=0.$

For the other direction we need to prove that for some real valued $\alpha$ and 
$\beta$ the operator
\begin{equation}
    X = \alpha C +\beta L
\end{equation}
has an entangled ground (lowest eigenvalue) state. This ground state can then be 
certified by the appropriate combination of $C$ and $L$, thus its entanglement  
can be detected.

Let us consider $\alpha = 1$ and $\beta = \lambda $, where $\lambda$ is a very 
small real number. We can assume that the operators $C_A$ and $C_B$ are diagonal 
in their respective local computational basis, so $C$ is diagonal in $\ket{kl}$, $k,l\in \{0,1\}$ 
with eigenvalues $\gamma_{kl}$. We need to distinguish two cases, depending on whether 
the operator $C$ has a degenerate ground state or not. 

{\it  First case:} Let us assume that $C$ has the unique ground state $\ket{00}$. 
Considering $\lambda L$ as a perturbation to $C$, the first order correction to the 
ground state is given by \cite{stingl}
\begin{equation}
\label{eq:firstordercorrection}
\ket{\psi_1}=\sum_{k\neq 0,l\neq 0} \frac{\bra{kl}L\ket{00}}{\gamma_{00}-\gamma_{kl}}\ket{kl}.
\end{equation}

We now prove the statement by contradiction, {\it i.e.}~we assume that the ground 
state of the operator $C+\lambda L$ is always a product state. For small values of 
$\lambda$, the ground state can, according to perturbation theory, be expanded as
\begin{equation}
\label{eq:expansion}
\ket{\psi(\lambda)} = \ket{00} + \lambda \ket{\psi_1} + \lambda^2\ket{\psi_2} + \dots.
\end{equation}
As the total state is normalized, $\ket{\psi_1}$ is orthogonal to $\ket{00}$. 

The first observation is that from the fact that $\ket{\psi(\lambda)}$ is a product 
state, it follows that $\ket{\psi_1}$ must also be orthogonal to all $\ket{kl}$, 
where $k,l>0$. For qubits, this only concerns the case $k=l=1$, but we formulate the
argument directly for arbitrary dimensions. This orthogonality can be seen as follows: 
Assume that $0 < f := \braket{kl}{\psi_1}$ for $k,l>0$, and consider the state 
$\ket{\varphi} = (\ket{00} + \ket{kl})/{\sqrt2}$. The state $\ket{\varphi}$ is 
entangled and it is known that for every product state $\ket{a,b}$ 
one has $\vert \braket{a,b}{\varphi}\vert^2\leq 1/2$ \cite{Guehne2009}.
For $\lambda=0$, we have that $\vert \braket{\psi(0)}{\varphi} \vert^2 = 1/2$ and
in addition  
\begin{equation}
\frac{\partial}{\partial\lambda} 
\vert \braket{\psi(\lambda)}{\varphi} \vert \big\vert_{\lambda=0}=f>0,
\end{equation}
so for small $\lambda$ the overlap obeys $|\braket{\psi(\lambda)}{\varphi}|^2 > 1/2$. 
Consequently $\ket{\psi(\lambda)}$ is entangled and we arrive at a contradiction.

Having established that $\ket{kl}$ for $k,l>0$ is orthogonal to the first order 
expansion vector $\ket{\psi_1}$ we can conclude from Eq.~(\ref{eq:firstordercorrection}) 
that
\begin{equation}
\label{eq:zero_constraint_terms}
\bra{11}L\ket{00}= \bra{00}L\ket{11}=0.
\end{equation}
Since $L=L_A\otimes L_B$, it follows that either $L_A$ or $L_B$ must be diagonal 
in the computational basis. This is the contradiction to the statement that 
$[C_x,L_x]\neq 0$ for $x\in \{A,B\}$.

{\it  Second case:}
Now we consider the case when $C$ is degenerate, in which case both the 
ground (lowest eigenvalue) state and the most excited (highest eigenvalue)
state must have two-fold degeneracy. This is because if only the ground state 
would be degenerate, the operator $-C$ could be used instead and the first 
case of the proof would apply.

Since we have the assumption that neither $C_A$ or $C_B$ commutes with $L_A$ or $L_B$, 
respectively, neither $C_A$ not $C_B$ can be proportional to the identity. It follows
that without loss of generality we can fix the degenerate ground subspace to be spanned by the two product 
vectors $\ket{00}$ and $\ket{11}$. Note that in this two-dimensional subspace,
$\ket{00}$ and $\ket{11}$ are the only product vectors.

The operator $C$ is disturbed by the operator $L$ and we want to characterize this
using degenerate perturbation theory \cite{stingl}. We define the projector $P=\ketbra{00}{00}+
\ketbra{11}{11}$ and, according to perturbation theory, we need to diagonalize 
the operator $PLP$. The ground state $\ket{\chi}$ of this operator is then the 
zeroth order of perturbation theory, that is in the limes $\lambda \searrow 0$
the ground state of the perturbed system approximates $\ket{\chi}$ arbitrarily 
well.

The vectors $\ket{00}$ and $\ket{11}$ cannot be eigenstates of $PLP$, as this would 
imply $\bra{11}L\ket{00}=\bra{00}L\ket{11}=0$ again. So, $\ket{\chi}$ must be entangled. 
But then there are no product states in its vicinity and for small $\lambda$ the operator
$C+\lambda L$ must have an entangled ground state.   
$\hfill \Box$

Note that Proposition 1 implies that even two jointly measurable observables can be 
used for entanglement detection. This is discussed in more detail in Section 3.

\subsection{Criteria for higher dimensions}

The question arises whether the same result is also true in higher dimensions. 
In a $2 \times 3$-dimensional system, a similar statement is true, except that 
in this case we need to ensure that the ground state of $C$ is non-degenerate.
For higher dimensions, we will present examples of $C=C_A \otimes C_B$ and 
$L=L_A \otimes L_B$ with $[C_A, C_B] \neq 0 \neq [L_A, L_B]$, where nevertheless
$C$ and $L$ cannot be used for entanglement detection.

\noindent
{\bf Proposition 2.}
{\it 
Consider a qubit-qutrit system and operators $C=C_A \otimes C_B$ and 
$L=L_A \otimes L_B$ where the ground state and the most excited state 
of $C$ are non-degenerate. Then $C$ and $L$ can be used for entanglement 
detection, if and only if $[C_A,L_A]\neq 0$ and $[C_B,L_B]\neq 0$.
}

{\it Proof.}
We assume again  that $C$ is diagonal in the computational basis and 
that the ground state is given by $\ket{00}$. Using the same methods 
as in the proof of Proposition 1, one can show that the first order 
correction to the ground state, $\ket{\psi_1}$, must be orthogonal 
to $\ket{00}$ and to all $\ket{kl}$, with $k,l>0$, i.e.~to $\ket{11}$ 
and $\ket{12}$. Similar orthogonality constraints hold for the 
corrections to the most excited state. Thus, the operator $L$ must 
have the following structure:
\begin{equation}
L=
\begin{pmatrix}\centerdot & \centerdot & \centerdot & \centerdot & 0 & 0\\
\centerdot & \centerdot & \centerdot & \centerdot & \centerdot & 0\\
\centerdot & \centerdot & \centerdot & \centerdot & \centerdot & \centerdot\\
\centerdot & \centerdot & \centerdot & \centerdot & \centerdot & \centerdot\\
0 & \centerdot & \centerdot & \centerdot & \centerdot & \centerdot\\
0 & 0 & \centerdot & \centerdot & \centerdot & \centerdot
\end{pmatrix}.
\end{equation}
Due to the product structure of $L$, this means that $L_A$ (or $L_B$) must 
be diagonal in the computational basis, too. This implies that it commutes 
with $C_A$ (or $C_B$), leading to a contradiction.
$\hfill \Box$

For the case of two qutrits, statements similar to Propositions 1 and 2 are not 
true. To show this, we present two Hermitean qutrit operators $L=L_A\otimes L_B$ 
and  $C=C_A \otimes C_B$ with $[C_A, C_B] \neq 0 \neq [L_A, L_B]$, where the operator 
$X=C+\lambda L$ does not have entangled ground or most excited states. This implies 
that all possible combinations of expectation values of $\exv{C}$ and $\exv{L}$ can 
origin from separable states and the pair of observables is useless for entanglement
detection. 

\begin{figure}[t]
    \centering
    \includegraphics[width=0.40\columnwidth]{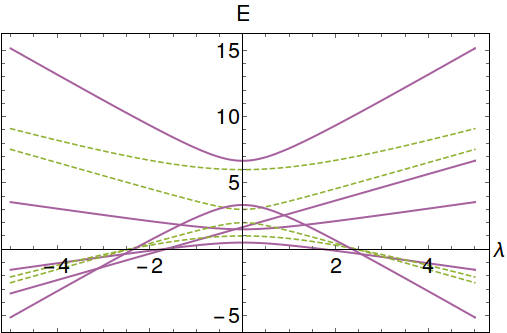}
    \hspace{0.05\columnwidth}
    \includegraphics[width=0.40\columnwidth]{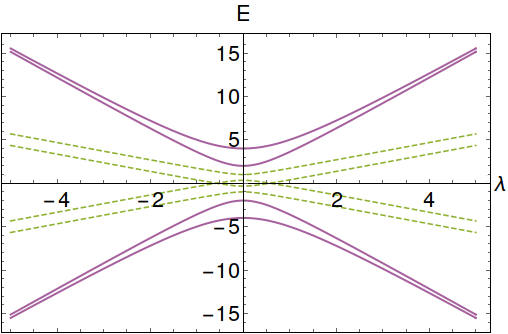}
    \caption{Left: Eigenvalues of the two-qutrit operator $X=C+\lambda L$ for 
    different values of $\lambda$ [see Eqs.~(\ref{eq:Cqutrit}, \ref{eq:Lqutrit}) 
    for the definitions of $C$ and $L$]. The dotted lines correspond to eigenvalues 
    of entangled eigenstates and the solid lines are depicting  eigenvalues of 
    product eigenstates. The lowest and highest eigenstates for the operator 
    $X$ always correspond to product states, yielding a counterexample for the 
    two-qutrit case. Right: Eigenvalues of the qubit-ququart operator $X=C+\lambda L$ for 
    different values of $\lambda$ [see Eqs.~(\ref{eq:Cququart}, \ref{eq:Lququart})]. 
    This gives a counterexample for the qubit-ququart case.} 
    \label{fig:groundstates}
\end{figure}

We take $C$ to be diagonal in the computational basis and $\ket{00}$ and $\ket{22}$ are 
its eigenstates corresponding to the lowest and highest eigenvalues. Requiring that the 
perturbed ground state remains to be a product state, we get conditions on the entries 
of the operator $L$. Similarly to Eq.~(\ref{eq:zero_constraint_terms}) for $k,l>0$ and 
$m,l<2$ the following  must hold:
\begin{equation}
\bra{kl}L\ket{00}=\bra{ml}L\ket{22}=0.
\end{equation}
Since $C$ and  $L$ are not diagonal in the same product basis, it follows that the Hermitean 
operators $L_A$ and $L_B$ can only have the structure
\begin{equation}
\label{eq:formsofL}
L_{A}=
\begin{pmatrix}\centerdot & 0 & 0\\
0 & \centerdot & \centerdot\\
0 & \centerdot & \centerdot
\end{pmatrix}\ \ \mbox{and }\ \ 
L_{B}=\begin{pmatrix}\centerdot & \centerdot & 0\\
\centerdot & \centerdot & 0\\
0 & 0 & \centerdot
\end{pmatrix},
 \end{equation}
or the other way round. In fact, choosing the diagonal 
matrices
\begin{equation}\label{eq:Cqutrit}
C_{A}=\begin{pmatrix}1 & 0 & 0\\
0 & 2 & 0\\
0 & 0 & 4
\end{pmatrix}\ \ \mbox{and }\ \ C_{B}=\begin{pmatrix}\frac{1}{2} & 0 & 0\\
0 & \frac{3}{2} & 0\\
0 & 0 & \frac{5}{3}
\end{pmatrix}
\end{equation}
and $L_A$ and $L_B$ to read
\begin{equation}\label{eq:Lqutrit}
L_{A}=\left(\begin{array}{ccc}
1 & 0 & 0\\
0 & 0 & 2\\
0 & 2 & 0
\end{array}\right)\ \ \mbox{and }\ \ L_{B}=\left(\begin{array}{ccc}
0 & \frac{1}{2} & 0\\
\frac{1}{2} & 0 & 0\\
0 & 0 & 1
\end{array}\right),
\end{equation}
one can explicitly calculate the ground and excited states of $X$ for all $\lambda$.
The result is displayed on the left side of Fig.~\ref{fig:groundstates}. It can be seen that the ground 
state and the most excited state always correspond to product states.

A similar counterexample can be found for the case of a qubit-ququart system: 
Choosing 
\begin{equation}\label{eq:Cququart}
C_{A}=\sigma_z\ \ \mbox{and }\ \ C_{B}=\begin{pmatrix}2 & 0 & 0 & 0\\
0 & \frac{1}{3} & 0 & 0\\
0 & 0 & -1 & 0\\
0 & 0 & 0 & 4
\end{pmatrix}
\end{equation}
and
\begin{equation}\label{eq:Lququart}
L_{A}=\sigma_x\ \ \mbox{and }\ \ L_{B}=\begin{pmatrix}3 & 0 & 0 & 0\\
0 & 0 & 1 & 0\\
0 & 1 & 0 & 0\\
0 & 0 & 0 & 3
\end{pmatrix},
\end{equation}
leads to the eigenvalue structure displayed on right side of Fig.~\ref{fig:groundstates}.

The presented counterexamples are surprising and relevant for the general construction 
of entanglement witnesses. In many situations, one tries to identify
two product observables, which are easy to measure. Then, in order to
obtain a strong witness, a typical recipe is to choose them ``locally 
anticommuting'', meaning that we have $\{C_X, L_X\}=0$ \cite{Toth2005}.
From our counterexamples we can conclude that in higher dimensions this
strategy may not always be successful.

\section{Discussion of the results of Ref.~\cite{Shahandeh2017}}
In this section we list the three main statements in 
Ref.~\cite{Shahandeh2017} and  give a detailed discussion 
of them. Two of the statements have already been corrected
in the erratum \cite{Erratum2017}, but it sometimes remains 
unclear, where precisely the mistake was made. Therefore, 
we present the issues in some detail. 

First, let us start with Theorem 1 of Ref.~\cite{Shahandeh2017} for which
it has already been pointed out  that it requires 
a revision \cite{Erratum2017}. The original version of  Theorem 1 in Ref.~\cite{Shahandeh2017}  
reads: {\it For a given constraint value $c$, the optimal
state $\sigma_{\rm opt; \mathcal X} \in \mathcal{S}_{\rm sep; \mathcal X}$ to the test 
operator $L$ is a pure state with ${\rm Tr}(C \sigma_{\rm opt; \mathcal X}) = c$.
} In the given context $\mathcal X=c$ and $\mathcal{S}_{\rm sep; \mathcal X}$ denotes 
all separable states obeying ${\rm Tr}(\varrho {C})=\mathcal X$ [see also 
Eq.~(\ref{eq:constrainedg}) in our introduction].

This statement is not correct, when looking for optimal states one cannot 
restrict the attention to pure states as shown by our counterexample already 
included in the erratum \cite{Erratum2017}: Consider a two-qubit system and 
take the operator $C = \ketbra{00}{00} - \ketbra{11}{11}.$ Then, for any 
given value $c \in [0,1]$ the state 
$\varrho_0 = [(1+c)/2] \ketbra{00}{00} + [(1-c)/2] \ketbra{11}{11}$ 
is in the plane described by ${\rm Tr}(\varrho {C})=c.$ 
If one takes the operator 
$L = \ketbra{00}{00} + \ketbra{11}{11}$
and maximizes its expectation value over pure states 
in $\mathcal{S}_{\rm sep; c}$ one finds by direct 
inspection that the value ${\rm Tr}( L \varrho_0 )$
is larger than the value for any pure state
in $\mathcal{S}_{\rm sep; c}$.

The error in the proof in Ref.~\cite{Shahandeh2017} is the following: The 
range of $\varrho_0$ is spanned by the vectors $\ket{00}$ and $\ket{11}$ 
and there are no other product vectors in the range. These product vectors 
do not lie in the plane characterized by ${\rm Tr}(\varrho {C})=c$. So, 
$\varrho_0$ constitutes already a counterexample to Lemma 3 in the Supplemental 
Material.\footnote{Note that the numbering of the Lemmata in the Supplemental
Material of Ref.~\cite{Shahandeh2017} differs from the arxiv version.} Going deeper, Lemma 2 in the Supplemental Material 
is  also not correct. This Lemma states that a vector $\ket{a^* b^*}$ occurs in 
some convex decomposition of $\sigma$, if $\ket{a^* b^*}$ is not in the kernel 
of $\sigma^{-1/2}.$ But in the proof it is assumed that 
$\sigma^{-1/2} \sigma^{1/2} = \eins$, which is not
correct if $\sigma$ is not of full rank. Note also that in the 
original version of this Lemma \cite{Cassinelli1997, Heinosaari2011}
the condition reads that $\ket{a^* b^*}$ should be in the range 
of $\sigma^{-1/2}$. 

Closing the discussion concerning Theorem 1, we add that in the erratum
\cite{Erratum2017} Theorem 1 is reformulated and states now correctly that 
the state $\sigma_{\rm opt; \mathcal X}$ is maximally of rank $2$. 

Second, we consider Theorem 2 in Ref.~\cite{Shahandeh2017}. This is ambiguous, 
as it reads: {\it The necessary condition for the separable operators $C$ and 
$L$ to detect entanglement [...] is that $[C, L] \neq 0$.}

The ambiguity comes from the notion  of a ``separable operator'', which 
is not  defined in Ref.~\cite{Shahandeh2017}. If one considers operators 
of the form $C = C_A \otimes C_B$ to be separable, then the statement is 
not correct. A counterexample for two qubits are the operators 
$C = \sigma_x \otimes \sigma_x$ and $L = \sigma_z \otimes \sigma_z$. 
They commute, but can be used for entanglement detection as we have seen 
in Section II. 

In the erratum Theorem 2 is reformulated and the commutator condition 
is replaced by the condition that the ``separable, positive operators''  
$C$ and $L$ ``are not diagonal in a common product basis''. Clearly, 
the statement is then correct even without a precise notion of  
``separable'' operators, and the positivity of the operators is not 
required for the conclusion.
But the necessary condition is then very weak and far from being sufficient, 
as we can learn from Proposition 1 in Section III. For instance, if 
$L = L_A \otimes L_B$ 
and  $C = C_A \otimes C_B$ with $[L_A, C_A]=0$ but $[L_B, C_B] \neq 0$,
the observables $C$ and $L$ are not diagonal in a common product basis, 
but they are useless for entanglement detection.

Finally, we consider Corollary 1, which has not been addressed in the 
erratum: {\it If $C = C_A \otimes C_B$ and $L = L_A \otimes L_B$
are product operators, then [in order to be able to detect entanglement] 
$C_Y$ and $L_Y$ ($Y = A,B$) must not be jointly measurable.}

This statement is not correct. This can be already inferred from 
Proposition 1 in Section III, but it is very instructive to discuss a 
counterexample. It can be found by considering noisy versions of Pauli 
measurements: For 
$E_\pm^C = (\eins \pm \sigma_x/\sqrt{2})/2$ and 
$E_\pm^L = (\eins \pm \sigma_z/\sqrt{2})/2$ it is well known that
all these effects are jointly measurable \cite{Heinosaari2011}. If one 
takes
$C_Y = E_+^C$ and $L_Y = E_+^L$ (for $Y = A,B$) and the 
corresponding $C$ and $L$ and restricts to the  hyperplane 
$c=0.6$ one can define an ultrafine entanglement witness
\begin{equation}
{W}_\text{UEW}=g_c \eins-{L},
\end{equation}
where a value of $g_c=0.5223$ can be obtained by a semidefinite program, 
using the separability criterion of the positivity of the partial transpose.
However, one can check that $\hat{W}_\text{UEW}$ attains a minimum of 
$-0.016$ over all states on the hyperplane defined by $c=0.6$. Thus,
the jointly measurable operators $C_Y$ and $L_Y$  are able 
to detect entanglement via an UEW. 

The motivation of this counterexample
comes from the fact that
\begin{equation}
{W} = \frac{9}{8}\eins - {C} - {L}
\end{equation}
is a standard entanglement witness: the expectation value for product 
states is non-negative, while ${W}$ has a negative eigenvalue. This 
implies that there must be values of $c$ for which an UEW with $L$ 
can detect entanglement.

The error in the proof in Ref.~\cite{Shahandeh2017} is the following. For $E_\pm^C$ 
and $E_\pm^L$ there is a generalized measurement  
with four outcomes on a qubit that allows to measure them jointly. 
This generalized measurement can be implemented as a projective 
measurement $G_Y$ on a four-dimensional space, and on this 
space $C_Y$ and $L_Y$ have a representation as commuting 
observables. Then, it is correct that any observed statistics of 
$G_A \otimes G_B$ can be mimicked by a separable state 
(see Eq.~3.1 in the Supplemental Material). But this does not imply 
that any statistics can be mimicked with a separable state on a 
$2 \times 2$ subsystem (of the total $4 \times 4$ space) only, 
as suggested in the proof of Corollary~1.  In other words, if 
one knows that a state acts on a certain  $2\times 2$ subspace 
only, one can detect entanglement with $G_A \otimes G_B$.

\section{Summary}

In summary, we have discussed the problem of entanglement detection, 
given the expectation values of two observables. We showed that the 
general problem can be tackled with the method of the Legendre 
transformation. We identified conditions for two product observables
to be  useful for entanglement detection. The conditions are necessary
and sufficient for two qubits, but the situation in the general case
is not clear. 

For further work, it would be interesting to clarify in a general 
setting which combinations of observables are useful for entanglement
detection. For instance, it would be interesting to identify three
product observables, for which each pair is useless for entanglement
detection, but the whole triple is useful. Such examples may then
be related to similar studies on joint measurability \cite{chris}
or state characterization \cite{teiko}.

\section{Acknowledgements}
We thank T. Heinosaari, M. Ringbauer, F. Shahandeh and R. Uola for 
discussions. We also thank T. G\"uhne and D. Wyderka for being quiet 
sometimes. This work was supported by the DFG and the ERC 
(Consolidator Grant 683107/TempoQ). M.G. acknowledges funding from 
the Gesellschaft der Freunde und F\"orderer der Universit\"at Siegen 
and N.W. acknowledges funding from the House of Young Talents
of the Universit\"at Siegen.


\end{document}